# An electronic warfare approach for deploying a software-based Wi-Fi jammer

Keshav Kaushik
*School of Computer Science, University of Petroleum and Energy Studies, Dehradun, India*
officialkeshavkaushik@gmail.com

Rahul Negi
*School of Computer Science, University of Petroleum and Energy Studies, Dehradun, India*
negir8106@gmail.com

Prabhav Dev
*School of Computer Science, University of Petroleum and Energy Studies, Dehradun, India*
prabhavdevgupta@gmail.com

***Abstract*— Some prominent instances have been centered on electronic warfare. For example, the American military has made significant investments in automation through UAV programs, only for competitors like the Iranians to create strategies to interfere with these systems. Iran managed to capture a top-secret U.S. surveil-lance drone by fooling it into descending in the incorrect place by jamming its control signals and providing it with bogus GPS data. In this paper, the authors have focused on the electronic warfare approach for deploying a software-based Wi-Fi jammer. The software-based Wi-Fi jammer can disconnect the targets using the DoS pursuit mode. The paper describes the unique methodology of how software can also be used for jamming wireless signals.**

## I. Introduction

There are now wireless networks everywhere. They are utilized all over the world in a variety of settings, including the home, the workplace, and even public spaces, to access the Internet and do business or conduct private affairs. Along with all the benefits of simplifying business and daily life, there are also certain hazards that need to be considered. Wireless network security issues have made it difficult to get into businesses, governments, and financial institutions. Because network managers are not completely coordinated when it comes to safeguarding wireless networks in a strong and dependable manner, the incidence of these assaults simply increases.

The notion of electronic warfare is very recent. Highly complicated procedures like GPS and Wi-Fi are conceivable because of communications and our way of life. Drones, Unmanned Aerial Vehicles (UAVs), and networked surveillance cameras may be attacked in a variety of methods, but their data transmission is frequently the most vulnerable. Instead of directly attacking these systems, electronic warfare prefers to interrupt and influence the data connections that these automated devices rely on. Such devices frequently stop working or return to anticipated and vulnerable default behavior when they don't have a consistent connection. A number of high-profile occurrences have included electronic warfare. For example, the US military [1] has substantially invested in automation through UAV programmes, only for opponents such as the Iranians to devise strategies to sabotage these technologies. Iran [2] was capable of capturing a top-secret US surveillance drone by jamming its command signal and supplying it bogus GPS data, causing it to crash in the wrong spot. In order to render American gadgets ineffective, the Russian military [3] has invested considerably in jamming and electronic warfare technologies. During a flight, Russia [4] even displayed the capacity to cripple a US battleship by knocking off the ship's energy.

The bulk of these sophisticated assaults are hardware-based and need the possession of gadgets that are either prohibited or prohibitively expensive. Thankfully, not all methods rely on hardware. The authors of this work focused on using an electronic warfare strategy to install a software-based Wi-Fi jammer.

## II. Related work

To render American technology worthless, the Russian military has also made significant investments in electronic warfare and jammer technology. Russia even showed off its capacity to take down a U.S. battleship while the ship was still powered up. Most of these potent assaults rely on hardware and call for tools that are either expensive or illegal to acquire. Thankfully, not all methods depend on hardware. In this paper [5], a Wi-fi DoS de-authentication attack is done for jamming the wireless networks. The present status of drone security is examined in this study [6], along with several Wi-Fi-enabled drone weaknesses. Two distinct types of commercially accessible drones were subjected to five various sorts of assaults, as well as the possibility of automating attacks. Attacks such as Denial of Service, De-authentication Methods, Man-in-the-Middle, Unauthorized Root Access, and Packet Impersonation attack are investigated in relation to the communications systems.

The issues with wireless network security are the subject of this paper [7]. There are several methods for finding current security flaws, including Kali Linux operating system. There are various processes involved in the general cracking process for wire-less networks. The effectiveness of the process depends on the speed with which the key can be cracked and the accuracy with which the password can be determined. The hardest part of cracking is this stage. The advantage [8] of this operating system is unquestionably the wide variety of various tools for vulnerability evaluation and penetration testing that are specifically created for ethical hacking.

The present status of drone security is examined in this study [9], along with several Wi-Fi-enabled drone weaknesses. Two distinct types of commercially accessible drones were subjected to five various sorts of assaults, as well as the possibility of automating attacks. Attacks such as Denial of Service, De-authentication Approach-es, Man-in-the-Middle, Unauthorized Root Access, and Packet Impersonation attack are looked into in relation to the communications systems.

Stochastic, sensing-based, and adaptive jammers are lessened by DeepWiFi [10]. A deep learning-based autoencoder is used in RF front end processing to extract frequency band characteristics. After that, a deep neural network is taught to accurately identify waveforms as Wi-Fi, jammer, or idle. With the help of channel labels, users may efficiently access busy or congested channels without interfering with legal Wi-Fi broadcasts that are verified by RF fingerprinting based on machine learning, leading to improved throughput.

To research and assess how well the radar system tracks targets under electronic attack conditions, an electronic warfare (EW) simulator is given [11]. The input portion of the EW simulator allows users to configure the distinctive characteristics of the radar threat, radar warning receivers, jammer, incident electromagnetic transmission, and simulation situation. The simulator estimates the transient and root-mean-squared sampling error of the range and angle monitoring system of radar while simulating various scenarios, including the received signal and its frequency, radar scope, and angle tracking extent.

Given how frequently Wi-Fi is used nowadays, it's crucial to demonstrate to students' practical ways to take use of Wi-Fi access points. The logic underlying WEP and WPA2 exploits has been around for a while. Offering students, the chance to use these theories in a real-world setting has not been simple, either. You will gain knowledge of the WEP vulnerability, the many hacking scenarios when WEP access points have linked client's vs no connected clients, and the WPA2 vulnerability in this paper [12]. Along with a WPA2 access point, you will also have the chance to hack up to four WEP access points.

The authors [13] of this research provide a summary of the photonics-based electronic warfare solutions that have been put up so far. We concentrate especially on the unique use of a recently suggested spectrum scanner. Here, we report on our most recent developments in tuning speed in addition to the effectiveness in terms of frequency, linearity, and responsiveness, all of which are comparable to the most cut-ting-edge online commerce systems. Additional enhancements are also considered regarding integrability and tuning efficiency.

## III. Working Methodology

In this section, the authors have discussed the working methodology followed in the implementation of results. A step-by-step process is shown in figure 1.

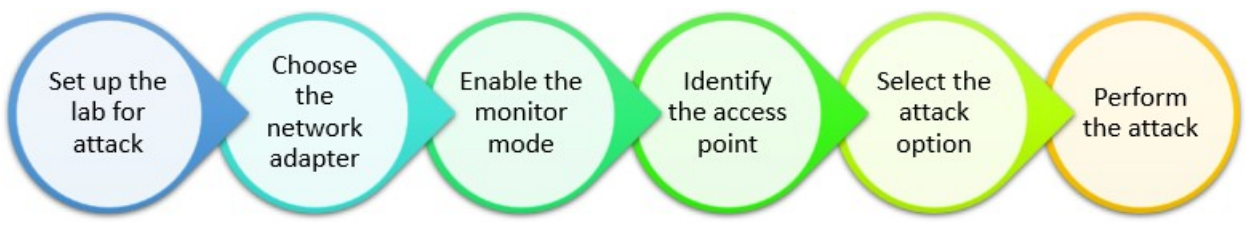

Fig 1: Methodology

- Lab setup- Firstly, the authors have set up the lab to perform the attack, in this step, the authors have set up kali Linux virtual machine and installed Airgeddon tool.
- Choose the network adapter - We will need a Wi-Fi adapter to perform Wi-Fi jamming on any network but only if attackers are running kali on a virtual machine. Wi-Fi adapter is this case is used to enable monitor mode in kali.
- Enable the monitor mode - Monitor mode allows us to capture packet which are not intended for us and to inject forged packets into targets network. We will be using Airgeddon tool to perform Wi-Fi jamming. It is a bash script which is used to perform audit on a network. To download Airgeddon tool in-to our kali machine we will clone it from the git repository using git clone command in our terminal. To run Airgeddon we change our present working directory to the directory in which Airgeddon is present. Now we will start Airgeddon by running the Airgeddon script "airgedddon.sh".
- Identify the access point - Now, Airgeddon checks for all the necessary repositories and updates, if all the repositories are present and it is up to date, then we proceed to next step. Now we are prompted with new Airgeddon interface where we are required to select our network interface which wlan0 which is wireless LAN or Wi-Fi in this case. After we continue, we are re-quired to put our network adapter in monitor mode which will allow us to capture packets or inject forged packets from targets network. So now we put interface in monitor mode by selecting option 2.
- Select the attack option - To continue with next step we select DoS attack mode which will provide us more options about DoS attack. Select DoS attack menu means option 4. Now to start with our attacks we first explore all the access points present in the range of our Wi-Fi adapter, it should be not-ed that our Wi-Fi adapter should be in monitor mode, so we select option 4. So, we get list of all access points within the range of our Wi-Fi adapter, now we have to select our targets access point along with the type of attack we are going to perform, so we will be doing disassociation attack which is type of DoS attack which will allow us to continuously disconnect all the devices that are connected to Wi-Fi.
- Perform the attack - Then we are prompted with a new interface where we are asked if we want to enable DoS pursuit mode, so enabling it will allow us to continuously execute our attack even if the target access point tries to de-fend from attack by switching channels. So, we enable it by selecting yes and press enter. Now the attack on the target access point starts, which continuously declines any connection request from various devices connected to that particular access point.

## IV. Results and Discussion

In this section, the authors mentioned about the results obtained. We have followed a methodology mentioned in the section 3. Figure 2 shows setting up the attacker ma-chine by installing the Airgeddon.

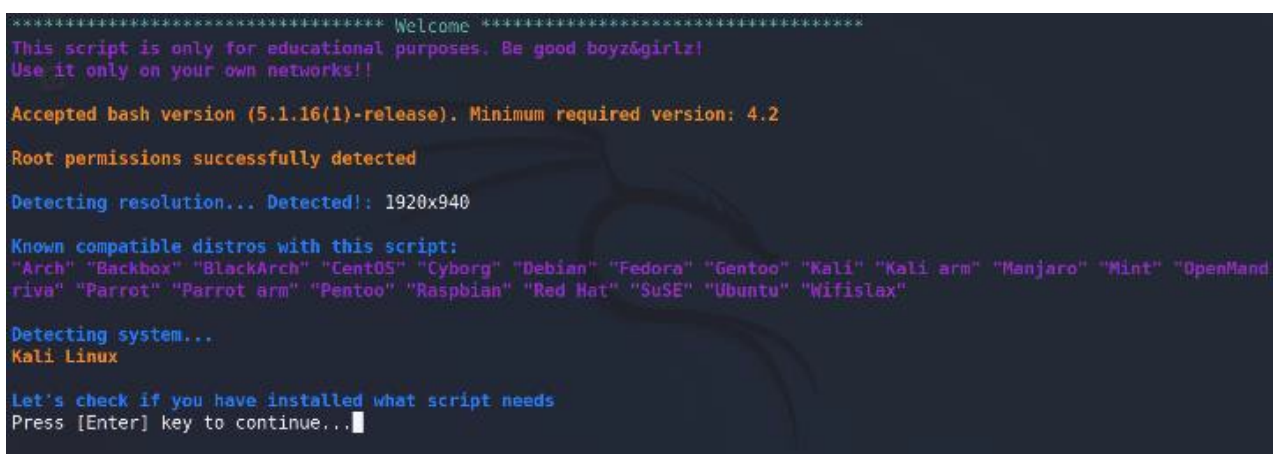

Fig 2: Setting Up

Now after we press enter to continue Airgeddon starts checking for any new updates or missing repositories. If no issues were found, then we get interface as shown in figure 3.

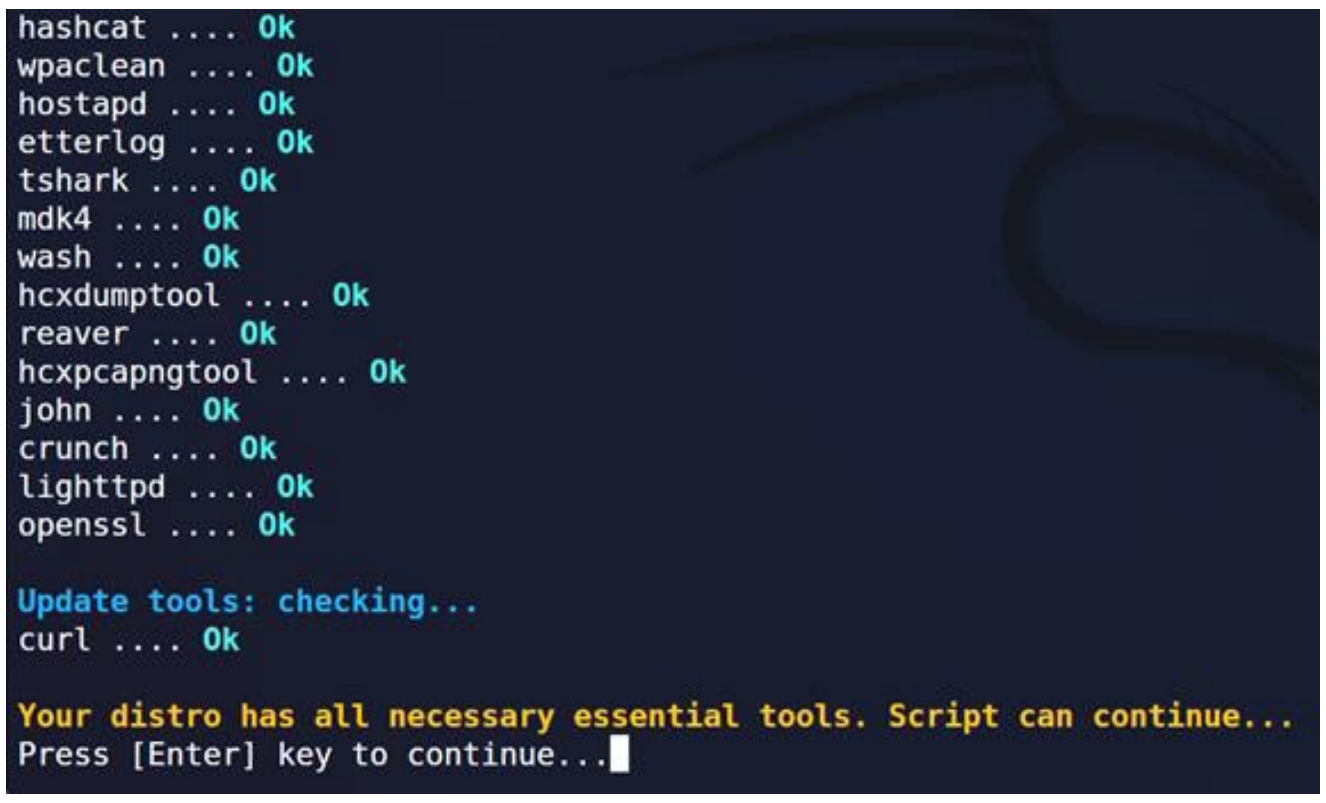

Fig 3: Checking for updates

We press enter to continue with our next step. Now in the next step we are displayed with all the network interfaces present in our system as shown in figure 4. Network interface in system in like a starting and endpoint of network connection between two or multiple devices. Here eth0 is for LAN connection and wlan0 is for wireless LAN or Wi-Fi. So, for our case we will work on wlan0 and select it by pressing 2.

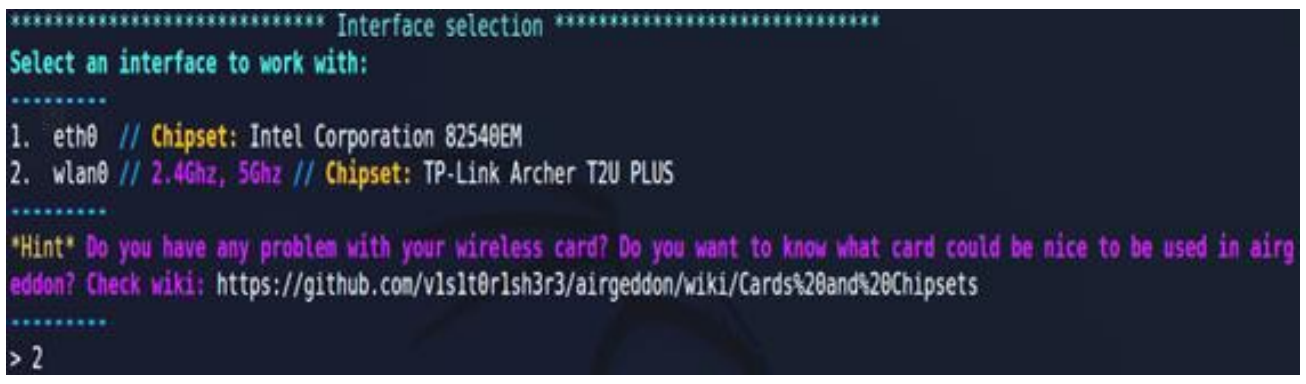

Fig 4: Network interfaces

After selecting our attack interface, we get our Airgeddon main menu which contains various options for different attack menus. Currently our interface is in managed mode, which is our default mode in which we are only allowed to receive packets which are meant for us. To inject packets, we switch our mode to “monitor mode”. In monitor mode we can inject forged packets and can also capture packets that are not meant for us.

Enabling monitor mode will allow attacker to inject forged packets in network and use them for various wireless attack. To enable monitor mode, we press 2 on our current Airgeddon interface as shown in figure 5.

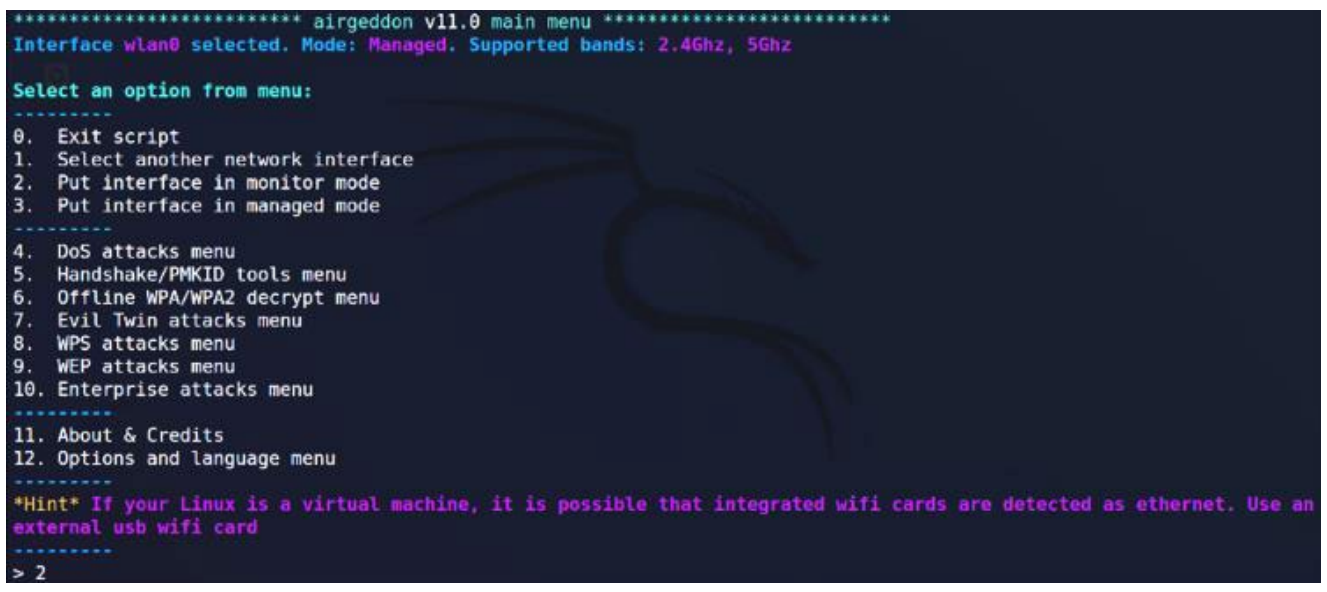

Fig 5: Monitor mode enabled

To continue with next step, we press enter. Now we are again prompted with the screen as shown in figure 5. This time we are supposed to select the attack type that we want to perform on the target network. Here we will select DoS attack menu which will enable us to perform a DoS attack which will flood the traffic on our target net-work forcing the router to decline the service requests from other devices.

To select DOS attack menu, we enter 4 and press enter to continue as shown in figure 6 and we get prompted with a new Airgeddon interface.

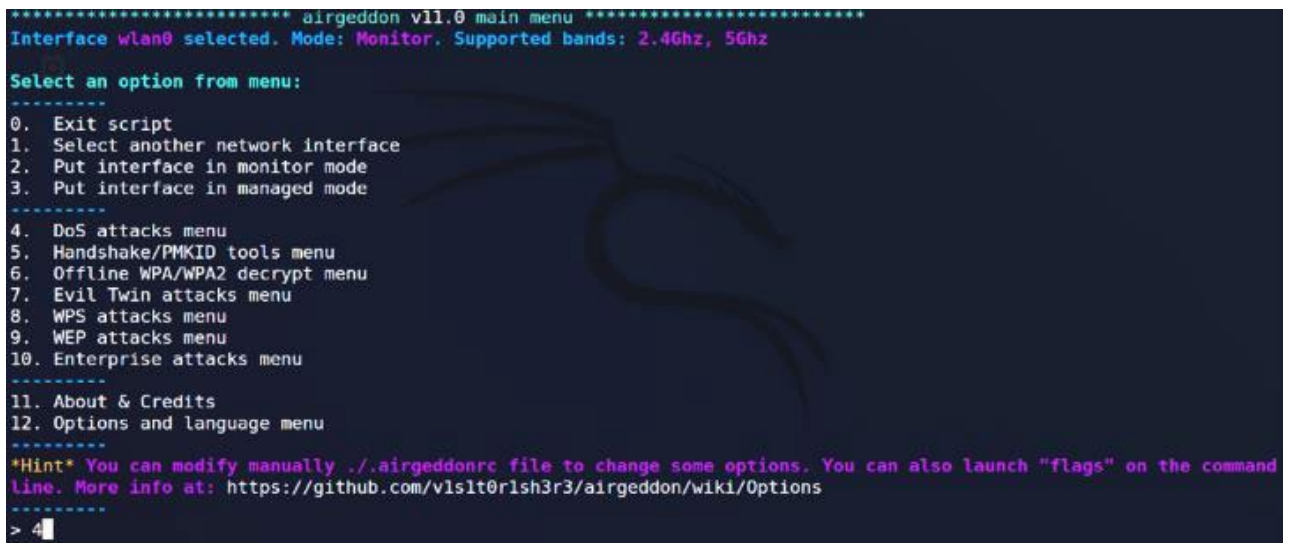

Fig 6: Attack type selection

After selecting the type of attack to be performed on target network, now we are supposed to scan for various targets which are present within range of our Wi-Fi adapter and we do this by selecting 4 on the current Airgeddon interface as shown in figure 7 and press enter to continue.

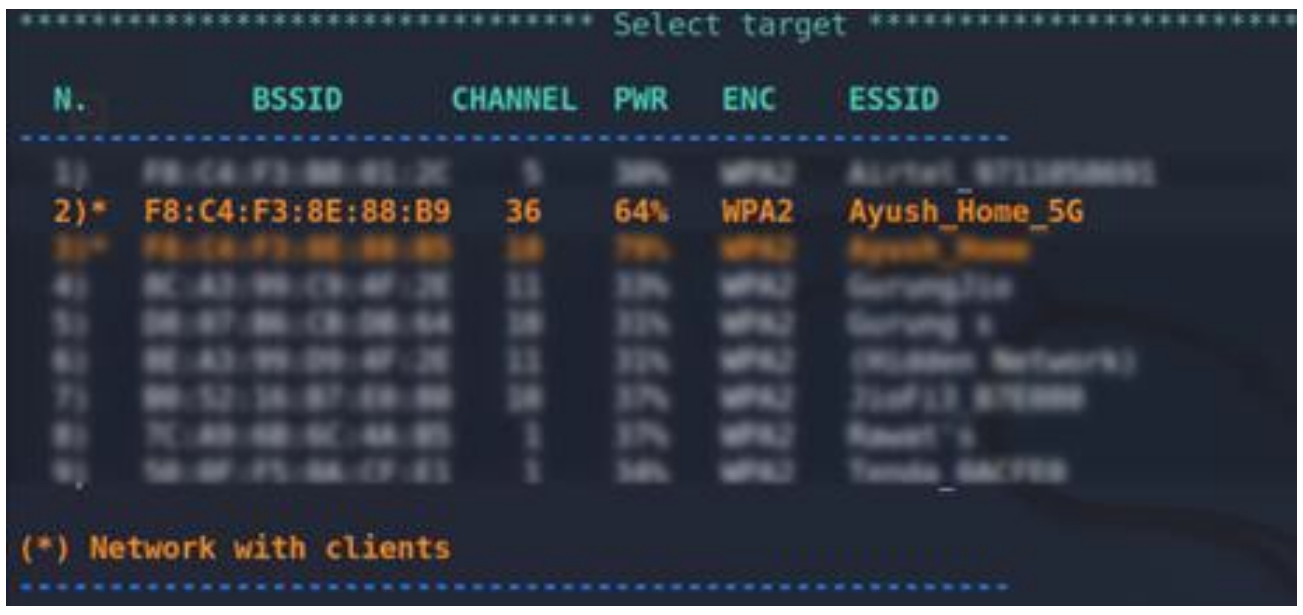

Fig 7: Available access points

Now as shown in figure 6 we get the list of all the Access Points present in the range of our network adapter. After few minutes of scanning, we get access points in our range we press CTRL-C to stop the scan. We get several access points after the scan. Now for this demonstration we will perform our attack on “Ayush_Home_5G”. In figure 7 we can see the mac address of this access point along with several other details such as encryption type which is WPA2 in for this access point. We select this access point by pressing 2 and then pressing enter to continue and we are prompted with new screen as shown in figure 8.

In our next step we are going to select our attack type that we want to perform on our target access point “Ayush_Home_5G” which is dissociation attack using mdk4 . This is a DoS attack where we will be flooding forged packets in our target network which will disconnect the devices connected on that network.

We select option 5 and press enter to continue as shown in figure 8.

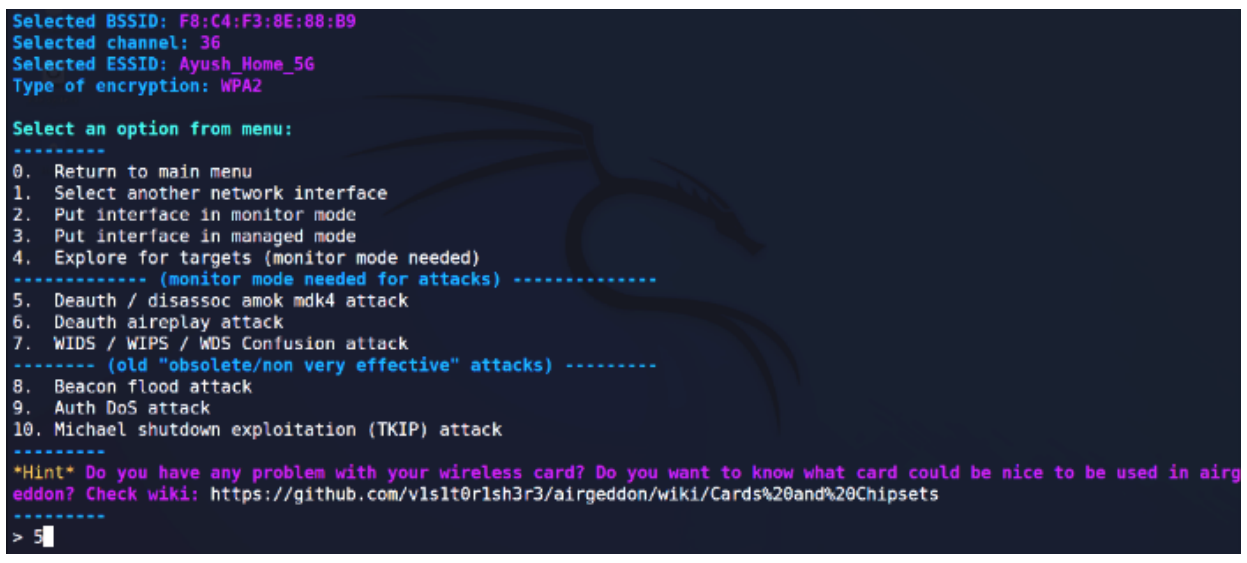

Fig 8: Selection of DOS attack type

Now in the next Airgeddon interface we are asked that whether we want to enable DoS pursuit mode or not. DOS pursuit mode enables us to launch attack continuously even if the target network tries to avoid the attack by switching its channel continuously. So, we select option "y" to enable DoS pursuit mode as shown in figure 9.

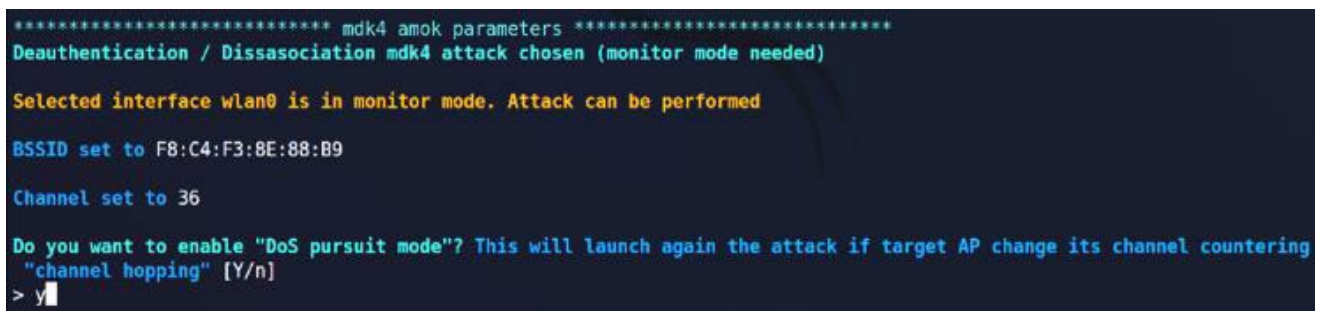

Fig 9: DOS pursuit mode

All the configurations required for the attack are now complete and we can launch the attack by pressing enter. We now press enter to start the attack. Now the attack begins, and we get a window which shows the progress of the attack. In figure 10 we see that various devices having different mac address are been disconnected from a particular mac address which is the mac address of our router present in "Ayush_Home_5G" network. This attack will continue disconnecting targets until it is stopped.

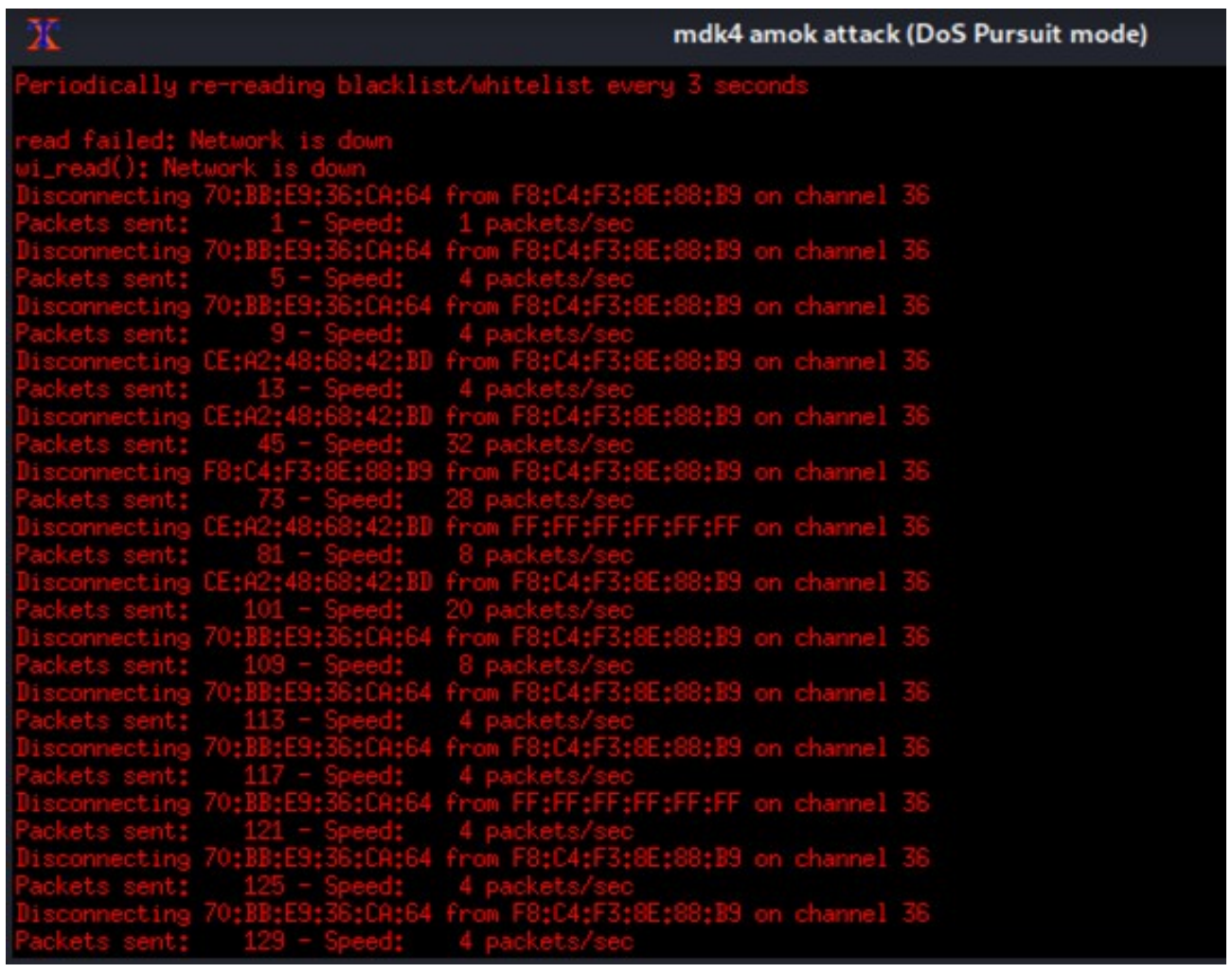

Fig 10: Disconnecting targets

## V. CONCLUSION

In this paper, the authors have demonstrated an electronic warfare approach to implement the software-based Wi-Fi jammer. Depends on how you use it and whether you have authority to audit the network you are targeting, this might be seen as a felony, just like any other DoS assault. If not, be warned that the router will save logs of the attack, which may be accessed to reveal the date and location of the attack, the MAC address involved, and other details that can be used to easily identify you through adjacent security cameras or cell tower data. The Upon considering all the facts the authors used a Wi-Fi adaptor for the virtual machine to perform Wi-Fi jamming and the tool used was Airgeddon. The authors performed the jamming on wlan0 port which is interface for wireless network. To perform the attack more effectively authors enabled the DOS pursuit mode which allow the attacker to continuously attack the target network even if the target network tries to avoid it by switching between different network channels. When the attack starts it will disconnect all the devices connected to Wi-Fi and will prevent them to connect again.

## REFERENCES

[1] "Video shows fighter jets launch swarm of tiny drones | Fox News." https://www.foxnews.com/tech/video-shows-fighter-jets-launch-swarm-of-tiny-drones (accessed May 26, 2022).

[2] "Exclusive: Iran hijacked US drone, says Iranian engineer - CSMonitor.com." https://www.csmonitor.com/World/Middle-East/2011/1215/Exclusive-Iran-hijacked-US-drone-says-Iranian-engineer (accessed May 26, 2022).

[3] "US Dim Mak point 2: Vulnerability to cyber/electronic warfare | The Manila Times." https://www.manilatimes.net/2017/01/12/opinion/analysis/us-dim-mak-point-2-vulnerability-cyberelectronic-warfare/306502 (accessed May 26, 2022).

[4] "Russia able to 'neutralise US warships', report claims | Daily Mail Online." https://www.dailymail.co.uk/news/article-4424320/Russia-able-neutralise-warships-report-claims.html (accessed May 26, 2022).

[5] "How to Hack Wi-Fi: Automating Wi-Fi Hacking with Besside-ng « Null Byte :: WonderHowTo." https://null-byte.wonderhowto.com/how-to/hack-wi-fi-automating-wi-fi-hacking-with-besside-ng-0176170/ (accessed Jun. 17, 2021).

[6] O. Westerlund and R. Asif, "Drone Hacking with Raspberry-Pi 3 and Wi-Fi Pineapple: Security and Privacy Threats for the Internet-of-Things," *2019 1st International Conference on Unmanned Vehicle Systems-Oman, UVS 2019*, Mar. 2019, doi: 10.1109/UVS.2019.8658279.

[7] P. Cisar and S. M. Cisar, "ETHICAL HACKING OF WIRELESS NETWORKS IN KALI LINUX ENVIRONMENT - ProQuest," *Annals of the Faculty of Engineering Hunedoara*, vol. 16, no. 3, pp. 181–186, Aug. 2018, Accessed: Oct. 11, 2022. [Online]. Available: https://www.proquest.com/docview/2125246008?pq-origsite=gscholar&fromopenview=true

[8] K. Kaushik, R. Tanwar, and A. K. Awasthi, "Security Tools," *Information Security and Optimization*, pp. 181–188, Nov. 2020, doi: 10.1201/9781003045854-13.

[9] A. Bhardwaj, K. Kaushik, and M. Kumar, "Taxonomy of Security Attacks on Internet of

Things," pp. 1–24, 2022, doi: 10.1007/978-981-19-1960-2_1.

[10] K. Davaslioglu, S. Soltani, T. Erpek, and Y. E. Sagduyu, "DeepWiFi: Cognitive WiFi with Deep Learning," *IEEE Trans Mob Comput*, vol. 20, no. 2, pp. 429–444, Feb. 2021, doi: 10.1109/TMC.2019.2949815.

[11] S. R. Park, I. Nam, and S. Noh, "Modeling and Simulation for the Investigation of Radar Responses to Electronic Attacks in Electronic Warfare Environments," *Security and Communication Networks*, vol. 2018, 2018, doi: 10.1155/2018/3580536.

[12] A. Ibrahim, "Building and Hacking an Exploitable Wi-Fi Environment for Your Classroom -- Even for Remote Participants!," pp. 1350–1350, Mar. 2021, doi: 10.1145/3408877.3432488.

[13] P. Ghelfi, F. Scotti, D. Onori, and A. Bogoni, "Photonics for Ultrawideband RF Spectral Analysis in Electronic Warfare Applications," *IEEE Journal of Selected Topics in Quantum Electronics*, vol. 25, no. 4, Jul. 2019, doi: 10.1109/JSTQE.2019.2902917.